\magnification\magstep1

\centerline{\bf The zero energy quantum information processing}

\medskip

\centerline{Miroljub Dugi\' c$^1$}

\medskip

\centerline{\it Department of Physics, Faculty of Science}

\centerline{\it P.O.Box 60, 34 000
Kragujevac, Yugoslavia}

\centerline{E-mail: dugic@knez.uis.kg.ac.yu}

\bigskip

{\bf Abstract:} In contradistinction with some plausible statements 
of the information theory, we point out the possibility of the zero
energy quantum information processing. Particularly, we investigate
the rate of the entanglement formation in the operation of the quantum 
"oracles" employing "quantum parallelism", and we obtain that the
relative maximum of the rate of the operation distinguishes the zero average 
energy of interaction in the composite system "input register + output
register". This result is reducible to neither of the previously
obtained bounds, and therefore represents a new bound for the nonorthogonal
state transformations in the quantum information processing.

\bigskip


\bigskip

{\bf 1. Introduction}

\bigskip

In the realm of computation, one of the central questions is "what
limits the laws of physics place on the power of computers?" [1].
Physically, this question refers to the minimum time needed for execution
of the logical operations, i.e. to the maximum rate of transformation of state 
of a physical system implementing the operation. From the fundamental point 
of view, this question tackles the yet-to-be-understood relation between 
the energy (of a system implementing the computation) on the one, and the
concept of information, on the other side. Eventually, answering this
question might shed new light on (e.g., might sharpen) the standard 
"paradoxes" of the quantum world [2].

Of special interest are the rates of the {\it reversible operations} (i.e.
of the reversible quantum state transformations). To this end, the two
bounds for the so-called "orthogonal transformations (OT)" are known;
by OT we mean a transformation of a (initial) state $\vert \Psi_i \rangle$
to a (final) state $\vert \Psi_f \rangle$, while $\langle \Psi_i \vert
\Psi_f \rangle = 0$. First, the minimum time needed for OT can be 
characterized in terms of the spread in energy, $\Delta \hat H$, of the
system implementing the transformation [3-7]. However, recently,
Margolus and Levitin [8, 9] have extended this result to show that a
quantum system with average energy $\langle \hat H \rangle$ takes time at
least $\tau = h / 4 \langle \hat H \rangle$ to evolve to an orthogonal 
state. In a sense, the second bound is more restrictive: a system with
zero average energy {\it cannot} perform a computation ever. This however
stems nothing about the nonorthogonal evolution which
is still of interest in quantum computation.

Actually, most of the efficient quantum algorithms [10-12] employ the
so-called quantum "oracles" (quantum "black boxes") employing the
"quantum parallelism" [13, 14]. These do not require orthogonality of
the initial and the final states of the composite quantum system "input
register + output register ($I+O$)". Rather, orthogonality of the final
states of the subsystems' (e.g. $O$'s) states is required, thus emphasizing 
a need for a new bound for the operation considered.

In this paper we show that, in general, the relative maximum of the rate of
the operation of the quantum "oracles" may point out the {\it zero average
energy} of interaction in the composite system $I+O$. More precisely: it
appears that the rate of the operation {\it cannot} be characterized in terms
of the average energy of the composite system as a whole. Rather, it can be 
characterized in terms of the average energy of interaction Hamiltonian, 
still pointing out the zero average energy of interaction.
Physically, in a sense, it means that as lower the average
energy, the higher the rate of the operation. This result is in obvious
contradistinction with the result of Margolus and Levitin [8, 9]. On the
other side, our result is neither reducible to the previously obtained bound
characterized in terms of the spread in energy [3-7], thus providing us 
with a new bound in the quantum information theory. Finally, the possibility
of the zero energy quantum information processing is somewhat counterintuitive
result, which, we believe might sharpen the distinction between the classical 
and the "quantum information".

\bigskip

{\bf 2. The quantum "oracle" operation}

\bigskip

It is worth emphasizing: we are concerned with the bounds characterizing the 
rate of (or, equivalently, the minimum time needed for) the {\it reversible}
transformations of a quantum system's states. Therefore, the bounds known
for the irreversible transformations are of no use here. Still, it is a
plausible statement that the information processing should be faster for a
system with higher (average) energy, even if--as it is the case in the 
reversible information processing--the system does not dissipate energy. 
This intuition of the classical information theory is justified by the
bound obtained by Margolus and Levitin [8, 9]. However, this bound refers to 
OT, and does not necessarily applies to the nonorthogonal evolutions.

The typical nonorthogonal transformations in the quantum computing theory are 
the operations of the quantum "oracles" {\it employing} 
{\it "quantum parallelism"}
[10, 14]. Actually, the operation considered is defined by the following
state transformation:
$$\vert \Psi_i \rangle_{IO} = \sum_x C_x \vert x \rangle_I \otimes
\vert 0 \rangle_O \to 
\vert \Psi_f \rangle_{IO} = \sum_x C_x \vert x \rangle_I \otimes
\vert f(x) \rangle_O, \eqno (1)$$

\noindent
where $\{\vert x \rangle_I\}$ represents the "computational basis" of the
input register, while $\vert 0 \rangle_O$ represents an initial state of the
output register; by "$f$" we denote the oracle transformation.

The point strongly to be emphasized is that the transformation (1) does {\it 
not} [10, 12] require the orthogonality $_{IO}\langle \Psi_i \vert \Psi_f
\rangle_{IO} = 0$. {\it Rather}, orthogonality for the subsystem's states
is required [10, 12]:
$$_O\langle f(x) \vert f(x') \rangle_O = 0, x \neq x'\eqno (2)$$

\noindent
for at least some pairs $(x, x')$, which, in turn, is neither necessary nor
a sufficient condition for the orthogonality $_{IO}\langle \Psi_i \vert
\Psi_f \rangle_{IO}$ to be fulfilled.

Physical implementation of the quantum oracles is an open question of the 
quantum computation theory. However (and in analogy with the quantum
measurement and the decoherence process [15-17]), it is well understood
that the implementation should rely on (at least indirect, or externally
controlled) {\it interaction} in the system $I+O$ as presented by the following
equality:
$$\vert \Psi_f \rangle_{IO} = \hat U(t) \vert \Psi_i \rangle_{IO}
\equiv \hat U(t) \sum_x C_x \vert x \rangle_I \vert 0 \rangle_O =
\sum_x C_x \vert x \rangle_I \vert f(x, t) \rangle_O, \eqno (3)$$

\noindent
where $\hat U(t)$ represents the unitary operator of evolution in time
(Schrodinger equation) for the combined system $I+O$; the index $t$
represents an instant of time, and we omit unnecessary symbol of the 
tensor product. Therefore, the operation (1) requires the orthogonality:
$$_O\langle f(x,t) \vert f(x',t) \rangle_O = 0, \eqno (4)$$

\noindent
which substituites the equality (2).

Therefore, {\it our task} in this paper reads: by the use of Eq. (4), we 
investigate the minimum time needed for  establishing of the entanglement 
present on the r.h.s. of both Eq. (1) and of Eq. (3).

\bigskip

{\bf 3. The optimal bound for the quantum oracle operation}

\bigskip

In this Section we derive the bound for the minimum time needed for the
execution of the transformation (1), i.e. (3), as distinguished by the
expression (4).

Actually, we consider the composite system "input register + output 
register ($I+O$)" defined by the Hamiltonian:
$$\hat H = \hat H_I + \hat H_O + \hat H_{int} \eqno (5)$$

\noindent
where the last term on the r.h.s. of (5) represents the interaction 
Hamiltonian. For simplicity, we introduce the following assumptions:
(i) $\partial \hat H / \partial t = 0$,
(ii)$[\hat H_I, \hat H_{int}] = 0$, $[\hat H_O, \hat H_{int}] = 0$, and
(iii) $\hat H_{int} = C \hat A_I \otimes \hat B_O$, where $\hat A_I$ and
$\hat B_O$ represent unspecified observables of the input and of the
output register, respectively, while the constant $C$ represents the
coupling constant.

\medskip

{\bf 3.1 Entanglement establishing}

\medskip

Given the above simplifications (i)-(iii), the unitary operator $\hat U(t)$
(cf. Eq. (3)) spectral form reads:
$$\hat U(t) = \sum_{x,i} \exp \{-\imath t (\epsilon_x + E_i +
C\gamma_{xi})/\hbar\} \hat P_{Ix} \otimes \hat \Pi_{Oi}.
\eqno (6)$$

The quantities in Eq. (6) are defined by the following spectral
forms: $\hat H_I = \sum_x \epsilon_x \hat P_{Ix}$, $\hat H_O =
\sum_i E_i \hat \Pi_{Oi}$, and $\hat H_{int} = C \sum_{x,i} \gamma_{xi}
\hat P_{Ix} \otimes \hat \Pi_{Oi}$; bearing in mind that $\hat A_I = 
\sum_x a_x \hat P_{Ix}$ and $\hat B_O = \sum_i b_i \hat \Pi_{Oi}$,
the eigenvalues $\gamma_{xi} = a_x b_i$.

From now on, we take the system's zero of energy at the
ground state by the exchange
$E_{xi} \to E_{xi} - {\bf E}_{\circ}$; $E_{xi} \equiv
\epsilon_x + E_i + C \gamma_{xi}$, ${\bf E}_{\circ}$
is the minimum energy of the composite system--which
Margolus and Levitin [8, 9] have used,
and Lloyd [1] as well. Then one easily
obtains for the output-register's states:
$$\vert f(x,t) \rangle_O = \sum_i
\exp\{-\imath t (\epsilon_x +
E_i + C \gamma_{xi} - {\bf E}_{\circ})/\hbar\}
\hat \Pi_{Oi} \vert 0 \rangle_O. \eqno 
(7)$$

Substitution of Eq. (7) into Eq. (4) directly gives:
$$D_{xx'}(t) \equiv _O\langle f(x,t) \vert f(x',t)
\rangle_O = \exp\{-\imath t (\epsilon_x - \epsilon_{x'})/
\hbar\} \times$$
$$\times \sum_i p_i \exp\{-\imath C t (a_x - a_{x'}) b_i/
\hbar\} = 0, \quad \sum_i p_i = 1, \eqno (8)$$

\noindent
where $p_i \equiv _O\langle 0 \vert \Pi_{Oi} \vert
0 \rangle_O$. The expression (8) is the condition of
the "orthogonal evolution" for the subsystem's ($O$'s) states
bearing explicit time dependence, while the ground
energy ${\bf E}_{\circ}$ does not appear in (8).

But this expression is already known from, e.g., the
decoherence theory [15-17]. Actually, one may 
write:
$$D_{xx'}(t) = \exp\{-\imath t (\epsilon_x - \epsilon_{x'})
/\hbar\} z_{xx'}(t), \eqno (9)$$

\noindent
where 
$$z_{xx'}(t) \equiv \sum_i p_i \exp\{-\imath C t (a_x - a_{x'})
b_i/ \hbar\} \eqno (10)$$

\noindent
represents the so-called {\it "correlation amplitude"}, 
which appears 
in the off-diagonal
elements of the (sub)system's ($O$'s) density matrix [15]:
$$\rho_{Oxx'}(t) = C_x C_{x'}^{\ast} z_{xx'}(t).$$

So, we could make direct application of the general
results of the decoherence theory. However, our aim is to
estimate the minimum time for which $D_{xx'}(t)$ 
may approach zero,
rather than calling for the qualitative limit of the
decoherence theory [15]:
$$\lim_{t \to \infty} \vert z_{xx'}(t)\vert = 0, \eqno (11)$$ 

\noindent
or equivalently $\lim_{t \to \infty}z_{xx'}(t) \to 0$.

So, {\it here}, we will use the inequality $\cos x \ge 1 -
(2/\pi) (x + \sin x)$, {\it valid only for} $x \ge 0$
[8, 9].  However, the use cannot be straightforward.

Actually, the exponent in the "correlation amplitude"
is proportional to:
$$(a_x - a_{x'}) b_i, \eqno (12)$$

\noindent
which need not be strictly positive. That is, for a
fixed term $a_x - a_{x'} > 0$, the expression Eq. (12)
can be both positive or negative, depending on the
eigenvalues $b_i$. For this reason, we will refer to the
general case of the eigenvalues of the observable $\hat B_O$,
$\{b_i, -\beta_j\}$, where both $b_i, \beta_j > 0$.

In general, Eq. (10) reads:
$$z_{xx'}(t) = z^{(1)}_{xx'}(t) + z_{xx'}^{(2)}(t),
\eqno(13a)$$

\noindent
where
$$z_{xx'}^{(1)} = \sum_i p_i \exp\{-\imath C t (a_x -
a_{x'}) b_i/\hbar\}, \eqno (13b)$$
$$z_{xx'}^{(2)} = \sum_j p'_j \exp\{\imath C t (a_x -
a_{x'}) \beta_j/\hbar\}, \eqno (13c)$$

\noindent
while $\sum_i p_i + \sum_j p'_j = 1$.
Now, since both $(a_x - a_{x'}) b_i >0$, $(a_x - a_{x'})
\beta_j >0, \forall{i, j}$, one may apply the above distinguished 
inequality.

The relaxed equality (4) (i.e. relaxed equality (11)) is
equivalent with $Re z_{xx'} \cong 0$ {\it and} $Im z_{xx'}
\cong 0$. Now, from Eq. (13a-c) it directly follows:
$$Re z_{xx'} = \sum_i p_i \cos [C (a_x - a_{x'}) b_i t / \hbar]
+ \sum_j p'_j \cos [C (a_x - a_{x'}) \beta_j t / \hbar],
\eqno (14)$$

\noindent
which after applying the above inequality gives:
$$Re z_{xx'} > 1 - {4 \over h} C (a_x - a_{x'}) 
(B_1 + B_2) t - {2 \over \pi} Im z_{xx'} -$$
$$- {4 \over \pi} \sum_i p_i \sin [C (a_x - a_{x'}) b_i
t/\hbar], \eqno (15)$$

\noindent
where $B_1 \equiv \sum_i p_i b_i$, and $B_2 \equiv \sum_j
p'_j \beta_j$. 

Now, since $\vert \sum_i p_i \sin 
[C (a_x - a_{x'}) b_i t/\hbar]\vert \le \sum_i p_i
\equiv \alpha < 1, \quad \forall{t}$, from Eq. (11)
and Eq. (15) it follows:
$$0 \cong Re z_{xx'} + {2 \over \pi} Im z_{xx'} >
1 - {4 \alpha\over \pi} - {4 \over h} C (a_x - a_{x'}) 
(B_1 + B_2) t. \eqno(16)$$

From (16) it is obvious that the
condition Eq. (4) cannot be fulfilled in the time
intervals shorter than $\tau_{xx'}$:
$$\tau_{xx'} > {(1 - 4 \alpha / \pi) h \over
4 C (a_x - a_{x'})(B_1 + B_2)}, \eqno (17)$$

\noindent
which is strictly positive for $\alpha < \pi/4$, and
which directly defines the optimal bound $\tau_{ent}$ as:
$$\tau_{ent} = sup \{\tau_{xx'}\}. \eqno (18)$$

The assumption $\alpha < \pi / 4$ is not very restrictive.
Actually, above, we have supposed that neither $\sum_i p_i \cong 1$,
nor $\sum_j p'_j \cong 1$, while the former is automatically 
satisfied with the condition $\alpha < \pi/4$.

\medskip

{\bf 3.2 Analysis of the results}

\medskip

The desired bound $\tau_{ent}$ is obviously determined by the 
minimum of the difference $a_x - a_{x'}$. This difference however
is virtually irrelevant. So, one may note that the bound Eq. (18)
can be {\it operationally} decreased by increase of the coupling 
constant $C$ and/or by the increase of the sum $B_1 + B_2$. As to 
the former, for certain quantum "hardware" [18], the coupling constant
$C$ can, at least in principle, be manipulated by experimenter. On
the other side, similarly--as it directly follows from the above
definitions of $B_1$ and $B_2$--by the choice of the initial
state of the output register, one could eventually increase the rate 
of the operation by the increase of the sum $B_1 + B_2$.

Bearing in mind the obvious equality:
$$\langle \hat H_{int}\rangle = \langle \hat A_I \rangle
\langle \hat B_O \rangle =  \langle \hat A_I \rangle
(B_1 - B_2), \eqno (19)$$

\noindent
one directly concludes that adding energy to the composite system
as a whole, does not necessarily increase the rate of the operation
considered. Rather, the increase of the rate of the operation is 
related to the average energy of interaction, $\langle \hat H_{int}
\rangle$. For instance, if $B_1 \neq 0$ while $B_2 = 0$, from Eq. (19)
it follows that the increase of $B_1$ coincides with the increase of
$\vert \langle \hat H_{int} \rangle\vert$, as well as with the
decrease of the bound Eq. (18). This observation is in accordance
with the bound obtained by Margolus and Levitin [8, 9]: the increase
of the average energy (of interaction) gives rise to the increase of
the rate of the operation.

However, for the general initial state of the output register, both 
$B_1 \neq 0$ and $B_2 \neq 0$. Then, e.g., for $B_1 > B_2$:
$$B_1 + B_2 = B_1 (1 + \kappa) \le 2B_1, \kappa \le 1, \eqno (20)$$

\noindent
which obviously determines the {\it relative maximum} of the rate
of the operation by the following equality:
$$B_1 = B_2, \kappa = 1, \eqno (21a)$$

\noindent
which, in turn (for $\langle \hat A_I \rangle \neq 0)$, is equivalent
with:
$$\langle \hat H_{int} \rangle = 0. \eqno (21b)$$

But this result is in {\it obvious contradistinction} with the
result of Margolus and Levitin [8, 9]. Actually, the expressions 
(21a,b) stem that, apart from the particular values of $B_1$
and $B_2$, the relative maximum of the rate of the operation {\it 
requires} (mathematically: implies) {\it the zero average energy of 
interaction}, $\langle \hat H_{int} \rangle = 0$.

\bigskip

{\bf 4. Discussion}

\bigskip

Intuitively, the speed of change of a system's state should be 
directly proportional to the average energy of the system. This
intuition is directly justified for the quantum "orthogonal 
transformations"
by the bound obtained by Margolus and Levitin [8, 9]. Naively, one
would expect this statement to be of relevance also for the 
nonorthogonal evolution. Actually, in the course of the orthogonal 
evolution, the system's state "passes" through a "set" of nonorthogonal 
states, thus making the nonorthogonal evolution faster than the
orthogonal evolution itself.

This intuition however is obviously incorrect for the cases studied.
In a sense, the expressions (21) state the opposite: as lower
difference $B_1 - B_2$ (i.e. as lower the average energy of interaction),
the faster the operation considered. Therefore, our the main result, 
Eq. (21), is in obvious contradistinction with the conclusion drawn
from the bound obtained by Margolus and Levitin [8, 9]: the {\it zero
average energy quantum information processing is possible} and, in
the sense of Eq. (21), {\it even preferable}. From the {\it operational}
point of view, the bound $\tau_{ent}$ can be decreased by manipulations of 
the interaction in the combined system $I+O$, as well as by the proper
{\it local operations} (e.g., the proper state preparations increasing
the sum $B_1 + B_2$) performed on the output register. 

As it can be easily shown, the increase of the sum $B_1 + B_2$ coincides
with the increase  of the spread in $\hat B_O$, $\Delta \hat B_O$,
i.e. with the increase in the spread $\Delta \hat H_{int}$. This
observation, however, cannot be interpreted as to suggest reducibility 
of the bound Eq. (18) onto the bound characterized in terms of the
spread in energy [3-7]--in the case studied, $\langle \hat H_{int}\rangle$. 
Actually, as it is rather apparent, the increase
in the spread $\Delta \hat H_{int}$ {\it does not} pose any restrictions 
on the average value $\langle \hat H_{int} \rangle$. Therefore, albeit
having a common element with the previously obtained bound [3-7], the
bound Eq. (17), i.e. Eq. (18), represents a new bound$^2$ in the quantum
information theory.

It cannot be overemphasized: the zero (average) energy quantum information
processing is possible, at least in principle. Moreover, the condition
$\langle \hat H_{int} \rangle = 0$ determines the relative maximum of the
operation considered. But this result challenges our classical intuition,
because it is commonly believed that the efficient information processing
presumes some "energy cost". In other words: one may wonder if "saving
energy" might allow the efficient information processing ever. 
Without ambition to give a definite answer to this question, we want to
stress: as long as the "energy cost" in the classical information processing 
(including the quantum-mechanical "orthogonal evolution") is surely
necessary, this {\it need not be the case with the quantum information 
processing}, such as the entanglement establishing. Actually, the
entanglement formation by no means represents acquiring the classical
information about the (sub)system(s). So, without further ado, we stress 
that Eq. (21) exhibits the peculiar aspect of the "quantum information"
(here: of the entanglement formation), so pointing to the necessity of its 
closer further investigation. To this end, the expression (21) might be
interpreted as to point to the boundary between the "classical information" 
and the "quantum information".

The roles of the two registers ($I$ and $O$) are by definition asymmetric, as 
obvious from Eq. (1) and Eq. (3). This asymmetry is apparent also in the bound 
Eq. (17), which is the reason we do not discuss in detail the role of the 
average value $\langle \hat A_I \rangle$. Having in mind the told in Section 
3, this discussion is really an easy task not significantly changing the
above conclusions.

Finally, the simplifications (i)-(iii) of Section 2 do not prove restrictive
for our considerations, as briefly discussed in Appendix I.

\bigskip

{\bf 5. Conclusion}

\bigskip

We show that the zero average energy quantum information processing is
possible. Concretely, we show that the entanglement establishing in
the course of operation of the quantum oracles employing "quantum parallelism",
distinguishes the zero average energy of interaction in the composite
system "input register + output register". More precisely: the zero
average energy of interaction proves to be optimal for execution of
the operation considered. This result challenges our classical intuition,
which plausibly stems a need for the "energy cost" in the information
processing. To this end, our result, which sets a new
bound for the nonorthogonal evolution in the quantum information processing,
might eventually be interpreted as to point to the boundary between the
"classical information" on the one, and of the "quantum information"--the
concept yet to be properly understood--on the other side.

\bigskip

{\bf Literature:}

\item{[1]}
S. Lloyd, "Ultimate physical limits to computation".
e-print arXive quant-ph 9908043

\item{[2]}
B. d'Espagnat, "Conceptual Foundations of Quantum Mechanics",
Benjamin, Reading, MA, 1971;
J.A. Wheeler and W.H. Zurek, "Quantum Theory and Measurement"
Princeton University Press, Princeton, 1982; 
Cvitanovi\' c et al, eds. "Quantum Chaos--Quantum Measurement",
Kluwer Academic Publishers, Dordrecht, 1992

\item{[3]}
S. Braunstein, C. Caves, G. Milburn,
Ann. Phys. {\bf 247}, 135 (1996)

\item{[4]}
L. Mandelstam, I. Tamm, J. Phys. (USSR) {\bf 9},
249 (1945)

\item{[5]}
A. Peres, "Quantum Theory: Concepts and Methods",
Kluwer Academic Publishers, Hingham, MA, 1995

\item{[6]}
P. Pfeifer, Phys. Rev. Lett. {\bf 70}, 3365
(1993)

\item{[7]}
L. Vaidman, Am. J. Phys. {\bf 60}, 182 (1992)

\item{[8]}
N. Margolus, L.B. Levitin, in {\it Phys.Comp96}, (eds.)
T. Toffoli, M. Biafore, J. Leao (NECSI, Boston)
1996

\item{[9]}
N. Margolus, L.B. Levitin, Physica D {\bf 120},
188 (1998)

\item{[10]}
D.R. Simon, SIAM J. Comp. {\bf 26}, 1474 (1997)

\item{[11]}
P.W. Shor, SIAM J. Comp. {\bf 26}, 1484 (1997);
see also Peter W. Shor, "Introduction to Quantum 
Algorithms", e-print arXive quant-ph 0005003

\item{[12]}
M. Ohya, N. Masuda, Open Systems and Information Dynamics, {\bf 7},
33 (2000)

\item{[13]}
A.M. Steane, Rep. Prog. Phys. {\bf 61}, 117 (1998)

\item{[14]}
D. Aharonov, in Annual Reviews of
Computational Phy\-si\-cs, vol. VI
(ed. Dietrich Stauffer, World
Scientific, Singapore, 1998);
J. Preskill, in "Introduction to Quantum 
Computation and Information", (eds.) H.K. Lo,
S. Popescu, J. Spiller, World Scientific,
Singapore, 1998

\item{[15]}
W.H. Zurek, Phys. Rev. D {\bf 26}, 1862 (1982);
Prog. Theor. Phys. {\bf 89}, 281 (1993)

\item{[16]}
D. Giulini, E. Joos, C. Kiefer, J. Kupsch,
I.-O. Stamatescu and H.D. Zeh, "Decoherence
and the Appearance of a Classical World in
Quantum Theory", Springer, Berlin, 1996

\item{[17]}
M. Dugi\' c, Physica Scripta {\bf 53}, 9 (1996);
{\bf 56}, 560 (1997)

\item{[18]}
A. Imamoglu et al, Phys. Rev. Lett. {\bf 83}, 
4204 (1999); D. Loss and D.P. DiVincenzo, Phys. Rev. 
A {\bf 57}, 120 (1998); G. Burkard at al, Phys. Rev. 
B {\bf 60}, 11404 (1999)

\item{[19]}
W.H. Zurek, Phys. Today {\bf 48}, 36 (1991)

\bigskip

{\bf Appendix I}

\bigskip

Relaxing the simplifications (i)-(iii) of Section 2 
does not lead to the
significant changes of the results obtained. This can be
seen by the use of the results of Dugi\' c [17], but for 
completeness, we will briefly outline the main 
points in this regard.

First, for a time dependent Hamiltonian, which is still a
"nondemolition observable", $[\hat H(t), \hat H(t')] = 0$,
the spectral form reads [17]:
$$\hat H = \sum_{x,i} \gamma_{xi}(t) \hat P_{Ix} \otimes
\hat \Pi_{Oi}. \eqno (I.1)$$

\noindent
This is a straightforward generalization of the cases studied.

Similarly, relaxing the exact compatibilities (cf. the 
point (ii) in
Section 2) leads to approximate separability--i.e., in
Eq. {\it (I.1)} appear the terms of the small norm--which does not
change the results concerning the "correlation amplitude"
$z_{xx'}(t)$ [15], and consequently concerning $D_{xx'}(t)$.

Finally, generalization of the form of the interaction
Hamiltonian (cf. point (iii) of Section 2) does not produce
any particular problem, as long as the Hamiltonian is of 
(at least approximately) separable kind, and also
a nondemolition 
observable. E.g., from $\hat H_{int} = \sum_k C_k \hat
A_{Ik} \otimes \hat B_{Ok}$, one obtains the term
$\sum_k C_k (a_{kx} - a_{kx'}) b_{ki}$, instead of the term
Eq. (12).

The changes of the results may occur [17] if the Hamiltonian
of the composite system is not of the separable kind and/or
not a "nondemolition observable".

For completeness, let us emphasize: a composite-system
observable is of the {\it separable kind} if it proves
{\it diagonalizable in a noncorrelated basis} of the Hilbert state
space of the composite system [17].

\vfill\eject

{\bf FOOTNOTES:}

$^1$E-mail: dugic@knez.uis.kg.ac.yu

$^2$This bound is of interest also for the decoherence theory, but 
it does not provide us with the order of the "decoherence time", $\tau_D$.
Actually, with inspection to Ref. [19], one may write--in terms
of our notation--that $\tau_D \propto (a_x - a_{x'})^2$, while--cf.
Eq. (17)-- $\tau_{ent} \propto a_x - a_{x'}$, which therefore stems
$\tau_D \gg \tau_{ent}$. This relation is in accordance with the
general results of the decoherence theory: the entanglement formation 
should precede the decoherence effect.

\end